\documentclass[conference]{IEEEtran}

\usepackage[firstpage=true]{background}
\usepackage{eso-pic}

\backgroundsetup{
  position=current page.south,
  angle=0,
  nodeanchor=south,
  vshift=1mm,
  hshift=-2.6mm,
  opacity=1,
  scale=3,
  color=blue!80!white,
  contents={\begin{varwidth}{\textwidth}\fontsize{3}{12}\selectfont 
  © 2023 IEEE. Personal use of this material is permitted. Permission from IEEE must be obtained for all other uses, in any current \\[-2.8em]
  or future media, including reprinting/republishing this material for advertising or promotional purposes, creating new collective \\[-2.8em]
  works, for resale or redistribution to servers or lists, or reuse of any copyrighted component of this work in other works. \\[-2.7em]
  This is the accepted version of the published paper with DOI: 10.1109/EuCNC/6GSummit58263.2023.10188332
  \end{varwidth}
  }
}

\usepackage{cite}
\usepackage{amsmath,amssymb,amsfonts}
\usepackage{graphicx}
\usepackage{textcomp}
\usepackage{xcolor}
\def\BibTeX{{\rm B\kern-.05em{\sc i\kern-.025em b}\kern-.08em
    T\kern-.1667em\lower.7ex\hbox{E}\kern-.125emX}}

\usepackage[utf8x]{inputenc}

\usepackage{xcolor}

\usepackage{amsmath, bm}
\usepackage{bigstrut}

\usepackage{subcaption}

\usepackage{cleveref}
\Crefname{figure}{Fig.}{Figs.}

\usepackage{siunitx}
\DeclareSIUnit\bit{b}

\usepackage[export]{adjustbox} % For frame around figure

\usepackage{pgf}
\usepackage{tikz}
\usetikzlibrary{shapes.geometric}
\usetikzlibrary{shapes.arrows}
\usetikzlibrary{arrows}
\usetikzlibrary{arrows.meta}

\usepackage{booktabs}
\usepackage{tabularx}
\newcolumntype{Y}{>{\centering\arraybackslash}X}
\usepackage{multirow}

\newcommand\clearrow{\global\let\rowmac\relax}
\clearrow

\usepackage[nogroupskip, nopostdot]{glossaries} % Add styling options
\glsdisablehyper 
\newacronym{ack}{ACK}{acknowledgement}
\newacronym{ampdu}{A-MPDU}{aggregated MAC protocol data unit}
\newacronym{amsdu}{A-MSDU}{aggregated \gls{msdu}}
\newacronym{aod}{AoD}{angle of departure}
\newacronym{aoi}{AoI}{age of information}
\newacronym{ap}{AP}{access point}
\newacronym{ar}{AR}{augmented reality}
\newacronym{a-bft}{A-BFT}{association beam forming training}
\newacronym{ber}{BER}{bit error ratio}
\newacronym{bft}{BFT}{beamforming training}
\newacronym{bf}{BF}{beamforming}
\newacronym{bi}{BI}{beacon interval}
\newacronym{bpc}{BPC}{bits per color}
\newacronym{brp}{BRP}{beam refinement period}
\newacronym{br}{BR}{beam refinement}
\newacronym{bss}{BSS}{basic service set}
\newacronym{bti}{BTI}{beacon transmission interval}
\newacronym{bt}{BT}{beam tracking}
\newacronym{cbap}{CBAP}{contention-based access period}
\newacronym{cots}{COTS}{consumer off-the-shelf}
\newacronym{cts}{CTS}{clear to send}
\newacronym{dmg}{DMG}{directional multi-gigabit}
\newacronym{dti}{DTI}{data transmission interval}
\newacronym{e2e}{E2E}{end-to-end}
\newacronym{evm}{EVM}{error vector magnitude}
\newacronym{fov}{FOV}{field of view}
\newacronym{fps}{FPS}{frames per second}
\newacronym{gi}{GI}{guard interval}
\newacronym{hmd}{HMD}{head-mounted display}
\newacronym{hpbw}{HPBW}{half power beamwidth}
\newacronym{ifs}{IFS}{inter-frame spacing}
\newacronym{imu}{IMU}{inertial measurement unit}
\newacronym{ip}{IP}{internet protocol}
\newacronym{kpi}{KPI}{key performance indicator}
\newacronym{lan}{LAN}{local area network}
\newacronym{ldpc}{LDPC}{low-density parity-check}
\newacronym{los}{LoS}{line of sight}
\newacronym{nlos}{NLoS}{non-line-of-sight}
\newacronym{olos}{OLoS}{obstructed \gls{los}}
\newacronym{m2p}{M2P}{motion-to-photon}
\newacronym{mcs}{MCS}{modulation and coding scheme}
\newacronym{mmw}{mmWave}{millimeter-wave}
\newacronym{mpc}{MPC}{multipath component}
\newacronym{mpdu}{MPDU}{MAC protocol data unit}
\newacronym{msdu}{MSDU}{MAC service data unit}
\newacronym{phy}{PHY}{physical layer}
\newacronym{ppdu}{PPDU}{PHY protocol data unit}
\newacronym{psdu}{PSDU}{PHY service data unit}
\newacronym{qoe}{QoE}{quality of experience}
\newacronym{aqoe}{AppQoE}{application quality of experience}
\newacronym{qos}{QoS}{quality of service}
\newacronym{rat}{RAT}{radio access technology}
\newacronym{rss}{RSS}{received singla strength}
\newacronym{rtp}{RTP}{real-time transport protocol}
\newacronym{rtxss}{\mbox{R-TXSS}}{responder transmit sector sweep}
\newacronym{rx}{RX}{receiver}
\newacronym{sls}{SLS}{sector-level sweep}
\newacronym{snr}{SNR}{signal-to-noise ratio}
\newacronym{sp}{SP}{service period}
\newacronym{ssw}{SSW}{sector sweep}
\newacronym{sta}{STA}{station}
\newacronym{stp}{STP}{setup message}
\newacronym{tof}{ToF}{time of flight}
\newacronym{tu}{TU}{time unit}
\newacronym{txss}{TXSS}{transmit sector sweep}
\newacronym{trn-t}{TRN-T}{training sequence}
\newacronym{trn-r}{TRN-R}{training sequence}
\newacronym{tru}{TRU}{training unit}
\newacronym{tx}{TX}{transmitter}
\newacronym{udp}{UDP}{user datagram protocol}
\newacronym{ue}{UE}{user equipment}
\newacronym{upa}{UPA}{uniform planar array}
\newacronym{uplink}{UL}{uplink}
\newacronym{ura}{URA}{uniform radial array}
\newacronym{vr}{VR}{virtual reality}
\newacronym{xr}{XR}{extended reality}
\newacronym{vr/ar}{VR/AR}{virtual/augmented reality}
\newacronym{thr}{THR}{throughput}
\newacronym{fbk}{FBK}{feedback}
\newacronym{me}{ME}{monitoring entity}
\newacronym{db}{DB}{database}
\newacronym{rl}{RL}{reinforcement learning}
\newacronym{soa}{SoA}{state-of-the-art}
\newacronym{mab}{MAB}{multi-armed bandit}
\newacronym{eirp}{EIRP}{equivalent isotropically radiated power}
\newacronym{cdf}{CDF}{cumulative distribution function}
\newacronym{ccdf}{CCDF}{complementary cumulative distribution function}

\newacronym{cir}{CIR}{channel impulse response}
\newacronym{ctf}{CTF}{channel transfer function}
\newacronym{mrc}{MRC}{maximum ratio combining}
\newacronym{hh}{HH}{horizontal-to-horizontal}
\newacronym{vv}{VV}{vertical-to-vertical}
\newacronym{agc}{AGC}{automatic gain control}
\newacronym{fspl}{FSPL}{free-space path loss}
\newacronym{ofdm}{OFDM}{orthogonal frequency-division multiplexing}
\newacronym{papr}{PAPR}{peak-to-average power ratio}
\newacronym{mimo}{MIMO}{multiple-input multiple-output}
\newacronym{fft}{FFT}{fast Fourier transform}
\newacronym{gpu}{GPU}{graphics processing unit}
\newacronym{de}{DE}{dominant eigenmode}

\newacronym{pdf}{PDF}{probability density function}

\newacronym{fdm}{FDM}{fully-digital \gls{mimo}}
\newacronym{det_full}{DET}{dominant eigenmode transmission}
\newacronym{det}{DET}{dominant eigenmode transmission}
\newacronym{sst}{SST}{spatial streaming transmission}
\newacronym{sm}{SM}{spatial multiplexing}
\newacronym{sage}{SAGE}{space alternating generalized expectation-maximization}
\newacronym{rf}{RF}{radio frequency}
\newacronym{eadf}{EADF}{effective aperture distribution function}
\newacronym{pmf}{PMF}{probability mass function}

\newacronym{isi}{ISI}{inter-symbol interference}
\newacronym{sc}{SC}{single carrier}
\newacronym{aoa}{AoA}{azimuth of arrival}
\newacronym{eoa}{EoA}{elevation of arrival}
\newacronym{gtd}{GTD}{geometrical theory of diffraction}

\usepackage{algorithm}
\usepackage{algpseudocode} 

\newcommand{\vspacebelowfig}{\vspace{-0.5 cm}}

\begin{document}

% \title{Millimeter-wave antenna placement for \\ enhancing extended reality QoE}
% \title{MANTRA: Millimeter-wave antenna array placement on head-mounted displays}

% \title{Millimeter-wave antenna arrays on head-mounted displays: how many are needed?}
% \title{Millimeter-wave for head-mounted displays: \\ how many antenna arrays and why?}
% \title{Millimeter-wave for head-mounted displays: \\ how adding antenna arrays enhances performance}
% \title{How many millimeter-wave antenna arrays should \\ a head-mounted display have?}

% \title{Millimeter-wave for head-mounted displays: \\ how many antenna arrays make sense?}

% \title{Dominant eigenmode performance for a head-mounted display with multiple mmWave arrays}

% \title{How many mmWave antenna arrays should a head-mounted display have and why?}
\title{Impact of Array Configuration on Head-Mounted Display Performance at mmWave Bands}
% \title{The impact of mmWave array configuration on the performance of head-mounted displays}

% \author{\IEEEauthorblockN{Alexander Marinšek}
% \IEEEauthorblockA{\textit{ESAT-WaveCore} \\
% \textit{KU Leuven}\\
% Ghent, Belgium \\
% 0000-0001-9696-5365}
% \and
% \IEEEauthorblockN{Xuesong Cai}
% \IEEEauthorblockA{\textit{Department of Electrical and Information Technology} \\
% \textit{Lund University}\\
% Lund, Sweden \\
% {0000-0001-7759-7448}}
% \and
% \IEEEauthorblockN{Lieven De Strycker}
% \IEEEauthorblockA{\textit{ESAT-WaveCore} \\
% \textit{KU Leuven}\\
% Ghent, Belgium \\
% 0000-0001-8172-9650}
% \and
% \IEEEauthorblockN{Fredrik Tufvesson}
% \IEEEauthorblockA{\textit{Department of Electrical and Information Technology} \\
% \textit{Lund University}\\
% Lund, Sweden \\
% 0000-0003-1072-0784}
% \and
% \IEEEauthorblockN{Liesbet Van der Perre}
% \IEEEauthorblockA{\textit{ESAT-WaveCore} \\
% \textit{KU Leuven}\\
% Ghent, Belgium \\
% 0000-0002-9158-9628}
% \and

\author{
\IEEEauthorblockN{Alexander Marinšek$^\ast$, Xuesong Cai$^\dagger$, Lieven De Strycker$^\ast$, Fredrik Tufvesson$^\dagger$, Liesbet Van der Perre$^\ast$}
\IEEEauthorblockA{
$^\ast$ESAT-WaveCore, KU Leuven, Ghent, Belgium \\
$^\dagger$Department of Electrical and Information Technology, Lund University, Lund, Sweden\\
\{alexander.marinsek, liesbet.vanderperre\}@kuleuven.be
}
}

\maketitle

\begin{abstract}
Immersing a user in life-like \gls{xr} scenery using a \gls{hmd} with a constrained form factor and hardware complexity requires remote rendering on a nearby edge server or computer. \Gls{mmw} communication technology can provide sufficient data rate for wireless \gls{xr} content transmission.
However, \gls{mmw} channels exhibit severe sparsity in the angular domain. This means that distributed antenna arrays are required to cover a larger angular area and to combat outage during \gls{hmd} rotation. At the same time, one would prefer fewer antenna elements/arrays for a lower complexity system. Therefore, it is important to evaluate the trade-off between the number of antenna arrays and the achievable performance to find a proper practical solution. This work presents indoor 28~GHz \gls{mmw} channel measurement data, collected during \gls{hmd} mobility, and studies the \gls{de} gain. \gls{de} gain is a significant factor in understanding system performance since \gls{mmw} channel sparsity and eigenmode imbalance often results in provisioning the majority of the available power to the \gls{de}. Moreover, it provides the upper performance bounds for widely\nobreakdash-adopted analog beamformers.
We propose 3 performance metrics -- gain trade\nobreakdash-off, gain volatility, and minimum service trade\nobreakdash-off -- for evaluating the performance of a multi-array \gls{hmd} and apply the metrics to indoor 28~GHz channel measurement data. Evaluation results indicate, that 3~arrays provide stable temporal channel gain. Adding a 4$^{th}$ array further increases channel capacity, while any additional arrays do not significantly increase physical layer performance.
\end{abstract}


\glsresetall

\section{Introduction}

% Stress problematic rotation, not movement

An \gls{xr} \gls{hmd} aims to immerse the user in a virtual environment or intertwine digital objects with the user's physical surroundings. An \gls{hmd} primarily focuses on deceiving the human visual system by means of high-resolution video data, requiring powerful \gls{gpu} rendering hardware. To avoid burdening the \gls{hmd} with additional processing components and batteries, rendering can take place at either an edge server \cite{FIROUZI2022101840} or a nearby computer, requiring transport medium data rates in excess of \SI{1}{Gbps} and more than \SI{10}{Gbps} for streaming unencoded video. \Glspl{hmd}, therefore, often rely on a tethered connection to the rendering machine. However, a cable will hinder user mobility and compromise \gls{xr} content fidelity -- the reason for needing to transfer video data in the first place. Wireless communication technologies 5G, 6G, and IEEE 802.11ad/ay can supply peak multi-Gbps data rates by leveraging the ample bandwidth availability in the \gls{mmw} and THz spectrum \cite{shafi_5g_2017,zhou_ieee_2018,THzMagazine}. Unfortunately, \gls{mmw} channels are known for their angular sparsity, i.e., having few distinct \glspl{mpc} \cite{cai_dynamic_2020, choi_measurement_2018}. Hence, an \gls{hmd}, equipped with widely\nobreakdash-adopted directional patch antennas (with a field of view less than $<90^{\circ}$ \cite{hoque_distinguishing_2014}) should feature several antenna arrays, placed along the \gls{hmd}'s outside perimeter, to provide sufficient angular coverage during rotation. In addition to misalignment, \gls{mmw} \glspl{mpc} are highly prone to blockage by other users and oneself \cite{gustafson_characterization_2012, vaha-savo_empirical_2020}. Adding antenna arrays can also mitigate the adverse effects of blockage since the broader angular coverage allows the \gls{hmd} to receive other \glspl{mpc} (reflections, unaffected by blockage).

\Gls{mmw} reception quality has recently been studied in the context of 5G NR mobile networks.
Handheld devices are subject to usage in adverse circumstances -- movement, orientation flipping, shadowing, and physical contact between its antennas and the user -- that can disturb the \gls{mmw} data link. Walking \SI{15}{\meter} from one end of a corridor to its far end will cause a degradation of around \SIrange{15}{20}{\decibel}, with additional $\pm$\SI{15}{\decibel} fluctuations depending on the specific user \cite{hejselbak_measured_2017}. Rotating the smartphone from vertical to horizontal orientation can further cause a two-fold change in the achieved performance, e.g., \cite{aggarwal_first_2019} notes a decrease in IEEE 802.11ad data rate by half upon orientation change. 
% Common causes are asymmetrical antenna radiation patterns and mismatched polarization. 
User body shadowing can effectively attenuate \SIrange{20}{25}{\decibel} in various handheld use cases \cite{zhao_user_2017}. A palm, holding the smartphone, will absorb up to \SI{15}{\decibel} signal strength, whereas a mere finger interacting with the handheld device can already disturb the \gls{mmw} data link gain by \SI{3}{\decibel} \cite{xu_radiation_2018}. All these issues can plague \gls{xr} \glspl{hmd} \gls{mmw} communications systems. Previous studies concerning \gls{mmw} channel measurements for \gls{xr} \glspl{hmd} have demonstrated, that high data rates are achievable in cluttered indoor scenarios even with limited \gls{mmw} link bandwidth \cite{gomes_mm_2018}, and that a device featuring a single phase\nobreakdash-shifted \gls{mmw} array will result in inappropriately volatile channel conditions for \gls{xr} video content streaming \cite{struye_opportunities_2022}. However, horn antennas and first\nobreakdash-generation \gls{cots} equipment, used in prior art, offer limited insight into how future \glspl{hmd}, employing \gls{mmw} technology might perform. In the work at hand, we utilize a multi-antenna-array \gls{mmw} channel sounder to assess channel conditions during small-scale \gls{hmd} movement due to head rotation, as well as large\nobreakdash-scale fading due to \gls{los} obstruction. \textbf{We pursue the answers to: 1)} What are the important physical layer performance metrics for configuring multiple \gls{mmw} antenna arrays on an \gls{hmd}? \textbf{2)} What is the minimal number of antenna arrays that still allows the \gls{hmd} to reap the benefits of high-bandwidth \gls{mmw} communications? Performance is evaluated based on the channel's \gls{de} gain. The reason for this is, that \gls{mmw} channels are sparse, and any additional eigenmodes, beyond the first, are of much lower quality. Hence, often the best strategy for applications with high data rate and somewhat looser reliability requirements, compared to vehicular and industrial applications, is allocating most if not all power to the dominant eigenmode. Leveraging on a dedicated measurement campaign, \textbf{our contributions are:}
\begin{itemize}
    \item \Gls{mmw} channel \gls{de} transmission performance is evaluated for a multi-array \gls{hmd} during rotation.
    \item Physical layer performance metrics, relevant for antenna array configuration on an \gls{hmd}, are proposed.
    % \item Channel conditions and trade-offs for a 1\nobreakdash--7~array \gls{hmd}, compared to a full 8-array system, are assessed.
    \item The profitability of using a rear headband on the \gls{hmd} as an antenna array mounting point is assessed.
\end{itemize} 

\Cref{sec:methodology} describes the experiment setup. The post\nobreakdash-processing procedures and proposed performance metrics are introduced in \Cref{sec:post_processing} and used in \Cref{sec:results} to evaluate \SI{28}{\giga\hertz} channel measurements. Lastly, \Cref{sec:conclusion} summarizes the work and outlines future research directions.

\section{Experiment setup}
\label{sec:methodology}

\subsection{Measurement equipment}

\begin{figure}[h]
    \centering
    \begin{center}               
        \resizebox{\columnwidth}{!}{\input{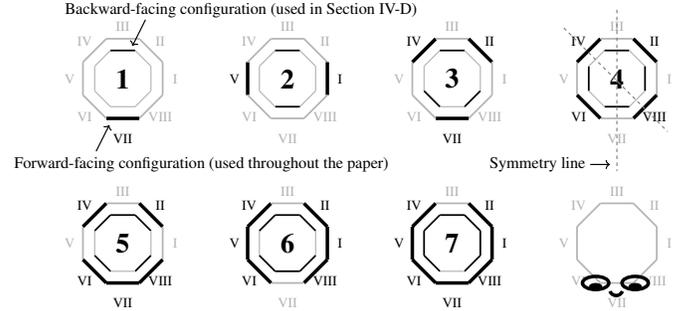}}
    \end{center}
    \caption{\gls{ue} top-down view, showing the forward- and backward-facing 1--7 panel configurations on the \gls{hmd}.}
    \label{fig:panel_configurations}
    \vspacebelowfig
\end{figure}

We collect \gls{cir} data using a switched array \gls{mimo} channel sounder \cite{sounder_paper}, custom-built by Lund University and SONY. The sounder operates at \SI{28}{\giga\hertz} with a \SI{768}{\mega\hertz} bandwidth (\SI{1.3}{\nano\second} time-delay resolution) and employs a Zadoff-Chu waveform. A top-down view of the \gls{ue} is illustrated in \Cref{fig:panel_configurations}. It consists of 8~planar patch antenna arrays, referred to as panels, offset by \SI{45}{\degree} in azimuth and featuring \SI{4}{}$\times$\SI{4}{} antennas each. \Cref{fig:panel_configurations} shows the studied forward-facing 1--7 panel configurations using the outer octagons. These are used throughout the work at hand, since we envision that future \gls{xr} \glspl{hmd} will prioritize both ergonomics and aesthetics; hence, they will not feature a rear headband by default. The inner octagons show the backward-facing configurations, evaluated only in \Cref{sec:rear_headband_results}. Conversely, the \gls{ap} features a single \SI{16}{}$\times$\SI{4}{} planar array. All antenna elements are dual-polarized, resulting in a \SI{256}{}$\times$\SI{128}{} channel matrix. Sampling is done at \SI{18.3}{\micro\second} per antenna element combination, which includes averaging over 4 sounding sequences to reduce measurement noise. The \gls{cir} snapshot sampling rate is  \SI{1}{\hertz} to allow additional time for memory writing. Rubidium clocks and additional preambles in the sounding waveform are used for accurate synchronization. \Cref{tab:system_parameters} summarizes the system parameters, while we refer the reader to \cite{tataria_275-295_2021} for further details on the channel sounder. \Cref{fig:measurement_environment_photo} includes a picture of the \gls{ue} and \gls{ap}. Note, that although we consider relatively large \gls{ue} antenna arrays (approx. \SI{2.5}{}$\times$\SI{2.5}{\centi\meter}), the learnings presented in this work are also applicable to smaller arrays with less elements since we study the trade-offs between a full 8\nobreakdash-panel setup and an \gls{hmd}\footnote{\gls{ue} refers strictly to the receiving octagonal antenna array structure, while \gls{hmd} is used in a broader context that emphasizes the studied use case.} with less antenna panels.

\begin{table}[h]
    \centering
    \vspace{-3mm}
    \caption{System parameters.}
    \vspace{-1mm}
    \begin{tabularx}{\columnwidth}{XY}
    % \begin{tabular}{l|c}
        \toprule    
        \textbf{Parameter} & \textbf{Value} \\
        \midrule
        Carrier frequency & \SI{28}{\giga\hertz} \\
        Bandwidth & \SI{768}{\mega\hertz}\\
        Time-delay resolution & \SI{1.3}{\nano\second}\\
        Max. observable time-delay & \SI{2.7}{\micro\second}\\
        Element combination sampling time & \SI{18.3}{\micro\second} \\
        \gls{cir} snapshot sampling frequency & \SI{1}{\hertz} \\
        Single-position measurement duration & \SI{33}{\second} \\
         & \\
        \gls{ap} array size & 4$\times$16\\
        \gls{ue} array size & 4$\times$4\\
        Number of \gls{ue} arrays (panels) & 8 \\ 
        \gls{cir} snapshot size & $256\times128\times2048$ \\
        \bottomrule
    \end{tabularx}
    \label{tab:system_parameters}
\end{table}

\subsection{Measurement environment}

\begin{figure}[h]
     \centering
     \begin{subfigure}[b]{0.99\columnwidth}
        \centering
        \begin{center}               
            \resizebox{0.68\columnwidth}{!}{\input{figure-conference/floorplan-stage1-horizontal}}
        \end{center}
        \vspace{-0.1 cm}
        \caption{Floor plan. Brown, cyan, gray, and magenta mark tables and the door, windows, metal whiteboards, and a projector screen, respectively. The room size is $6\times$\SI{9.15}{\meter}.}
        \label{fig:measurement_environment_floorplan}
     \end{subfigure}
     \hfill
     \begin{subfigure}[b]{0.99\columnwidth}
        \centering
        \begin{center}               
            \resizebox{0.99\columnwidth}{!}{\input{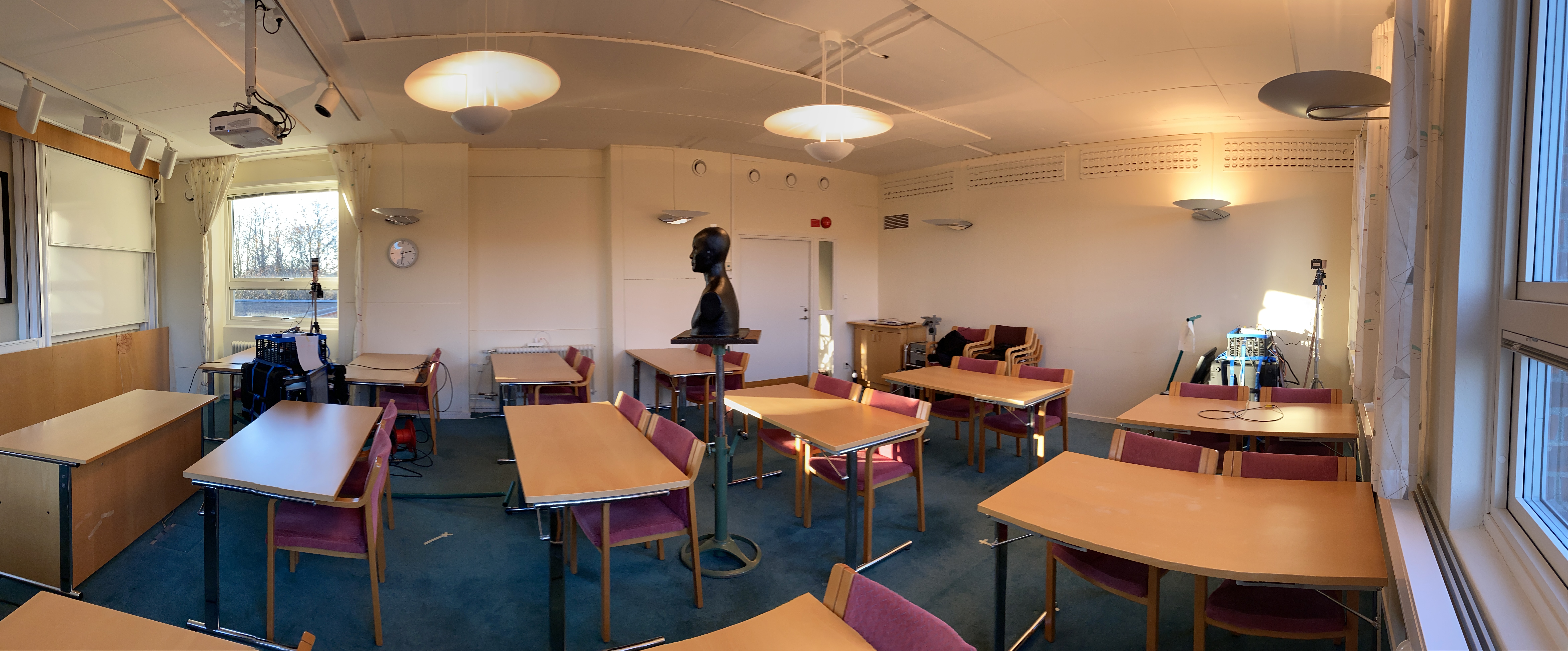}}
        \end{center}
        \vspace{-0.2 cm}
        \caption{Setup for UE$_6$, \gls{nlos}. Superimposed: mean difference between the phantom's attenuation and that of 6~human subjects (in \SI{}{\decibel}).}
        \label{fig:measurement_environment_photo}
     \end{subfigure}
    \caption{Conference room measurement environment.}
    \label{fig:measurement_environment}
    \vspacebelowfig
\end{figure}

Measurements were carried out at 11 discrete random positions in a conference room, depicted in \Cref{fig:measurement_environment}. The initial \gls{ue} orientation is randomly selected for each position and is shown by the red line at each \gls{ue} position in \Cref{fig:measurement_environment_floorplan}. This orientation serves as the starting point for the mobility pattern, described in \Cref{sec:mobility_pattern}. Channel sounding is repeated at each position for \gls{nlos} conditions, where a fiberglass water-filled human phantom \cite{gustafson_characterization_2012} represents a person obstructing the \gls{los}. Since the phantom was initially designed for \SI{60}{\giga\hertz} experiments, we first characterized it against 6 volunteers at \SI{28}{\giga\hertz}. The difference in attenuation between the phantom and the human subjects ranges from \SI{-0.6}{} to \SI{2.1}{\decibel}, hence, the phantom is suited for \SI{28}{\giga\hertz} experiments. The environment remains static during \gls{cir} sampling, except for \gls{ue} mobility, detailed in the upcoming subsection.

\subsection{Mobility pattern}
\label{sec:mobility_pattern}

\begin{figure}[h]
    \centering
    \includegraphics[width=0.92\columnwidth]{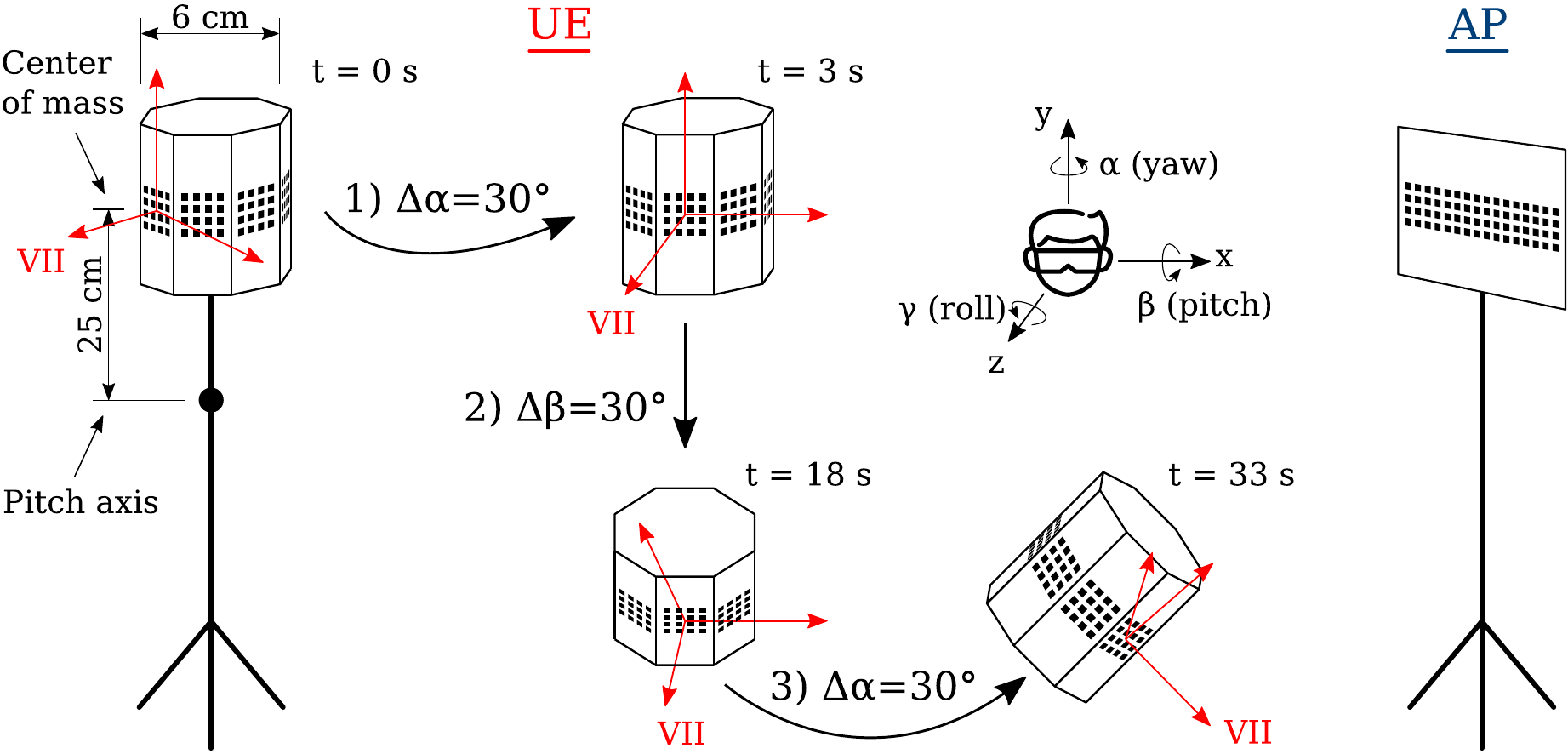}
    \caption{\gls{ue} mobility pattern in extrinsic Euler rotations.}
    \label{fig:sounder_and_mobility_sequence}
    \vspacebelowfig
\end{figure}

A \gls{ue} mobility pattern, outlined in \Cref{fig:sounder_and_mobility_sequence}, has been designed to mimic different possible \gls{hmd} rotations during usage. Initially, the \gls{ue} is standing upright with panel VII oriented in the direction shown in \Cref{fig:measurement_environment_floorplan}. Recall from \Cref{fig:panel_configurations} that panel VII is oriented in the \gls{hmd} looking direction. A $\Delta\alpha=30^\circ$  leftward (positive) yaw rotation is executed first, followed by pitching $\Delta\beta=30^\circ$ towards the ground, and concluded by a second $\Delta\alpha=30^\circ$ leftward rotation in yaw, while pitched downwards. All rotations are of extrinsic $xyz$ Euler type. Movement is executed by manually rotating the tripod's head, on which the \gls{ue} is mounted. Antenna velocity is kept below \SI{1}{\centi\meter\per\second}, thus, less than one carrier wavelength per second. Considering the \SI{25}{\centi\meter} distance (marked on the left side of \Cref{fig:sounder_and_mobility_sequence}) between the center of rotation and the octagonal \gls{ue}'s center of mass -- representing the offset between an \gls{hmd} and the human cervical vertebrae \cite{kunin_rotation_2007} -- the $30$\nobreakdash-$30$\nobreakdash-$30^\circ$ pattern requires $3$\nobreakdash-$15$\nobreakdash-$15$ \SI{}{\second} or \SI{33}{\second} in total. We refer to the resulting 33~\gls{cir} matrices as \textit{one measurement}. Note, that the second and third rotation displaces the \gls{ue}'s center of mass by approximately \SI{13}{} and \SI{6.5}{\centi\meter}, correspondingly, which make it lean beyond the phantom during parts of \gls{nlos} measurements.

\section{Post-processing procedures}
\label{sec:post_processing}

In this work, we evaluate the \gls{de} gain, i.e., the gain of single-spatial-stream transmission with channel information at both \gls{ap} and \gls{ue}, per \gls{ofdm} subcarrier, for different \gls{ue} antenna panel configurations. As previously mentioned, the combination of high data rate and loose reliability requirements, combined with channel sparsity, in practice results in the majority of power being allocated to the \gls{de}. For example, when applying waterfilling \cite{molisch_wireless_book}. Furthermore, when evaluated over the entire bandwidth, the \gls{de} transmission scheme's gain provides both the upper gain bound for analog beamformers, often found in \gls{mmw} \gls{cots} devices, and the average gain of a joint spectral-spatial \gls{ofdm}\nobreakdash-\gls{mimo} precoder, commonly found in lower-frequency cellular and IEEE 802.11 networks. Note that fully-digital \gls{mimo} and hybrid beamforming can in practice reach higher channel capacity in the absence of a dominant signal component, while the latter also manages to maintain relatively high energy efficiency \cite{qi_energy_2022}. Both fully-digital \gls{mimo} and hybrid beamforming are beyond the scope of this work. For clarity, \Cref{tab:index_overview} lists the relevant indices in this section.

\begin{table}[h]
    \centering
    \vspace{-3mm}
    \caption{Index overview.}
    \vspace{-1mm}
    % \begin{tabular}{c|l}
    %     \toprule
    %     \textbf{Index} & \textbf{Meaning} \\
    %     \midrule
    %     p & Number of \gls{ue} panels \\
    %     u & \gls{ue} position \\
    %     s & \gls{los}/\gls{nlos} scenario \\
    %     i & Channel snapshot index \\
    %     k & Subcarrier index \\
    %     \bottomrule
    % \end{tabular}
    \begin{tabularx}{\columnwidth}{>{\hsize=.35\hsize}YX|>{\hsize=.35\hsize}YX}
        \toprule
        \textbf{Index} & \textbf{Meaning} & \textbf{Index} & \textbf{Meaning} \\
        \midrule
        p & Number of \gls{ue} panels & i & Channel snapshot index \\
        u & \gls{ue} position & k & Subcarrier index \\
        s & \gls{los}/\gls{nlos} scenario & & \\
        \bottomrule
    \end{tabularx}
    \label{tab:index_overview}
\end{table}

\subsection{De-noising and dominant eigenvalue extraction}
\label{sec:denoising}
We first de\nobreakdash-noise the \gls{cir} to minimize the amount of noise that propagates into the frequency domain upon Fourier transformation (sum over \gls{cir} samples), enabling us to observe lower \gls{de} gain, e.g., for a single array configuration in \gls{nlos} conditions, without it sinking below the noise floor. Let $\bm{H}(\tau)$ denote the 2-dimensional \gls{mimo} \gls{cir} at delay $\tau$ and let $\lambda_1(\tau) \geq ... \geq \lambda_n(\tau)$ be the squared singular values of $\bm{H}(\tau)$ (eigenvalues of $\bm{H}\bm{H}^H$). We apply the de-noising procedure outlined in \Cref{alg:denoising}. First, we select $\lambda_1(\tau)$ values at time delays far exceeding any reasonable or observed \gls{mpc} delays, i.e., $\tau \in (1.35, 2.7)$ \SI{}{\micro\second} (second half of the \gls{cir}). The result is a vector of the largest eigenvalues of 1024 noisy square matrices, denoted as $\bm{\lambda}_{\text{noise}}$. The values have a Tracy-Widom $\mathcal{T}\mathcal{W}_2$ distribution \cite{chiani_distribution}, and, due to the limited \gls{cir} dimensions, we use the 95$^{th}$ percentile as the noise threshold, instead of applying, for example, the Marčenko-Pastur law. Next, we apply the eigenvalue threshold $\Lambda$ for setting noisy \gls{cir} samples to 0. Finally, all \gls{cir} entries exceeding the $\tau_{\text{max}} = 105$ \SI{}{\nano\second} delay spread, corresponding to \SI{31.5}{\meter} or the longest path traveled by a 2$^{nd}$ order reflection, are suppressed to 0. This leaves us with at most 80 non-zero \gls{cir} samples. We estimate the potential gain loss by considering the worst-case scenario, where 64 of the 80 windowed samples (\SI{105}{\nano\second}) represent at least 2$^{nd}$ order reflections with a gain just below the threshold $\Lambda$, and that \SI{18}{\decibel} separates the gain of the \gls{los} and the reflected \glspl{mpc} -- \SI{8}{\decibel} for path loss and \SI{2}{}$\times$\SI{5}{\decibel} for reflections. Then the calculated \gls{de} gain after de-noising would be \SI{3}{\decibel} too low. Contrarily, removing 64/80 noisy \gls{cir} samples from the summation in the Fourier transform allows us to study \SI{18}{\decibel} smaller eigenmodes in the frequency domain. 

\begin{algorithm}
\caption{\gls{cir} de-noising}
\begin{algorithmic}[1]
\algrenewcommand\algorithmicindent{0.5em} %Reduce indentation to fit comments
\Procedure{De-noise}{$\bm{H}$}
	\State $\bm{\lambda}_{\text{noise}} =\lambda_{1}(\tau), \text{ for } \tau \in (1.35, 2.7)$ \SI{}{\micro\second} \Comment{Evaluate noise}
	\State $\Lambda = P_{95}( \bm{\lambda}_{\text{noise}} )$ \Comment{Get threshold}
	\State $\bm{H}(\tau) \gets \bm{0}, \text{ where } \bm{\lambda}_{\text{noise}} < \Lambda$ \Comment{Below threshold}
	\State $\bm{H}(\tau) \gets \bm{0}, \text{ for } \tau > \tau_{\text{max}} $ \Comment{Out of delay spread}
\EndProcedure
\end{algorithmic}
\label{alg:denoising}
\end{algorithm}

We convert each \gls{cir} to its \gls{ctf} using the \gls{fft}, before extracting the dominant eigenmode gain per subcarrier. We use a 2048-point \gls{fft} for convenience since the \gls{cir} has 2048 time-delay samples. All further mention of eigenvalues/modes refers explicitly to the frequency domain. We drop the index $_1$ in view of brevity and denote the $k$-th tone's dominant eigenmode as $\lambda[u,s,i,k]$, where $k \in [0,2047]$ (\SIrange{27.616}{28.384}{\giga\hertz}), while $u$, $s$, and $i$ represent the \gls{ue} position, scenario (\gls{los}/\gls{nlos}), and channel snapshot index, correspondingly (see \Cref{tab:index_overview}).

\subsection{Performance metrics}

% This section proposes processing procedures for evaluating the performance of a 1--7~panel \gls{hmd}, based on $\lambda[u,s,i,k]$.
This section proposes processing procedures for evaluating the performance of a 1--7~panel \gls{hmd}, based on $\lambda[u,s,i,k]$, where $u,s,i,k$ are listed in \Cref{tab:index_overview}.

\subsubsection{Gain trade-off}
The first metric is the \gls{de} gain trade-off between a full 8-panel \gls{hmd} and a p-panel \gls{hmd}. It allows us to identify whether changing the panel count results in merely a gain difference proportional to beamforming gain or if there are other driving forces behind the difference, such as diversity gain. The trade-off is derived as follows:
\begin{equation}
    \label{eq:gain_tradeoff}
    \overline{\Delta\lambda_p}[u,s,i] =  \frac{1}{K} \sum_{k=0}^{K-1} \frac{\lambda_{p}[u,s,i,k]}{\lambda_{8}[u,s,i,k]},
\end{equation}
where $\lambda_p[u,s,i,k]$ and $\lambda_8[u,s,i,k]$ are the \glspl{de} of a p\nobreakdash-panel and an 8\nobreakdash-panel \gls{hmd}. An average over 2048 subcarriers for each channel snapshot constitutes the final value. 

\subsubsection{Gain volatility}
An important performance metric for \gls{xr} applications is wireless link stability, since it determines how much video stream buffer is required, how often the resolution is adapted, and whether there is a need for radio access technology switching. Channel gain spread and persistence are evaluated through the gain's standard deviation and autocorrelation, respectively. The latter is derived according to:
\begin{equation}
    r_p[u,s] = \frac
    { \sum\limits_{i=0}^{I-2} ( \overline{\lambda_{p}}[u,s,i]\!-\!\overline{\overline{\lambda_{p}}}[u,s] ) ( \overline{\lambda_{p}}[u,s,i\!+\!1]\!-\!\overline{\overline{\lambda_{p}}}[u,s] ) }
    { \sum\limits_{i=0}^{I-1} ( \overline{\lambda_{p}}[u,s,i] - \overline{\overline{\lambda_{p}}}[u,s] )^2 },
\end{equation}
where the number of snapshots per measurement is $I=33$, $\overline{\lambda_{p}}[u,s,i]$ is the average \gls{de} over all subcarriers, and $\overline{\overline{\lambda_{p}}}[u,s]$ represents its mean value. The evaluation is carried out independently at each measurement position and for \gls{los}/\gls{nlos} scenarios. A high standard deviation will indicate a large \gls{de} gain spread during the course of one measurement, while a low autocorrelation is the result of frequent changes in gain over time. Observing both high standard deviation and low autocorrelation, in combination with low average gain, would signal a potentially unstable \gls{xr} service since there are large variations in the already adverse channel conditions.

\subsubsection{Minimal service and capacity trade-off}
We obtain the minimal service trade-off by comparing the 3$^{rd}$ percentile \gls{de} gain for a p-panel configuration against that of an 8-panel \gls{hmd}. We use the 3$^{rd}$ percentile based on the \SI{97}{\percent} link reliability constraint, corresponding to the largest allowed error ratio for satisfactory encoded video quality \cite{nightingale_subjective_2013}. The metric shows how much channel capacity is lost due to employing less panels in the most adverse conditions, without considering \SI{3}{\percent} of the lowest recorded gains. Assuming \SI{}{SNR}$_p >> 1$, the minimal service trade-off is calculated as follows \cite{molisch_wireless_book}:
\begin{equation}
     \Delta C_{p,97} \; \approx \;  \left| \log_2 \left( \frac{P_3(\bm{\overline{\lambda_{p}}})}{P_3(\bm{\overline{\lambda_{8}}})} \right) \right|, 
\end{equation}
where $\bm{\overline{\lambda_{p}}}$ represents a vector of the \glspl{de} at each position, scenario, and snapshot ($u$, $s$, $i$), averaged over all 2048 subcarriers ($k$). 
Furthermore, we derive the channel capacity trade-off for each individual measurement, which allows us to evaluate how much capacity is wasted due to employing less than 8~panels. It is calculated according to:
% \begin{equation}
%      \overline{\Delta C_p} =  \frac{1}{K} \sum_{k=0}^{K-1} \left( C_p[k] - C_8[k] \right) \; \approx \; \frac{1}{K} \sum_{k=0}^{K-1} log_2 \left( \frac{\lambda_{p}[k]}{\lambda_{8}[k]} \right) 
% \end{equation}
\begin{equation}
     \overline{\Delta C_p}[u,s,i] \; \approx \; \left| \frac{1}{K} \sum_{k=0}^{K-1} \log_2 \left( \frac{\lambda_{p}[u,s,i,k]}{\lambda_{8}[u,s,i,k]} \right) \right|,
\end{equation}
% \begin{equation}
% \begin{aligned}
%     \overline{\Delta C_p} 
%     &=\overline{log_2 \left( \frac{\lambda_{p}[u,s,i,k]}{\lambda_{8}[u,s,i,k]} \right)} \\
%     &\approx \; \frac{1}{K} \sum_{k=0}^{K-1} log_2 \left( \frac{\lambda_{p}[u,s,i,k]}{\lambda_{8}[u,s,i,k]} \right) \\
% \end{aligned}
% \end{equation}
% where $C_p[k]$ is the p-panel capacity per subcarrier and $\overline{\Delta C_p}$ represents the mean capacity trade-off, compared to an 8\nobreakdash-panel \gls{hmd}.
where $\overline{\Delta C_p}$ represents the mean channel capacity trade-off over all 2048 subcarriers.

\subsection{Profit from a rear headband}
Lastly, we evaluate the benefit of fitting a rear headband to the \gls{hmd} and placing an antenna panel on it by assessing the overall gain difference between the default, forward-facing, panel configuration and its rotated counterpart. The latter always employs the backward-facing panel III, wheres the former excludes it (see \Cref{fig:panel_configurations}). We use a similar approach to \Cref{eq:gain_tradeoff} and evaluate the rear headband benefits using:
\begin{equation}
    \overline{\Delta\lambda_p^{rh}}[u,s,i] =  \frac{1}{K} \sum_{k=0}^{K-1} \frac{\lambda_{p,\text{back}}[u,s,i,k]}{\lambda_{p,\text{front}}[u,s,i,k]},
\end{equation}
where $\lambda_{p,\text{front}}[u,s,i,k]$ and $\lambda_{p,\text{back}}[u,s,i,k]$ represent the forward- and backward-facing p-panel configuration \gls{de} gains.
% , correspondingly (outer and inner octagons in \Cref{fig:panel_configurations}). 
% Values greater than 1 (positive in \SI{}{\decibel}reception quality has recently) indicate a bias towards backward-facing configurations.

\section{Results}
\label{sec:results}
% This section introduces the processing procedures for extracting 4 \glspl{kpi} from $\lambda[u,s,i,k]$ and assessing the performance of different \gls{hmd} panel configurations, before evaluating the profitability of an antenna panel on the \gls{hmd}'s rear headband.
% This section proposes the relevant \gls{de} performance metrics for \gls{mmw} array placement on an \gls{hmd}, and applies them to assess the behaviour of 1--7~array \gls{hmd}.

\subsection{Gain trade-off and impact of mobility}
\label{sec:results_general}

\begin{figure}[h]
    \centering
    \resizebox{0.98\columnwidth}{!}{\input{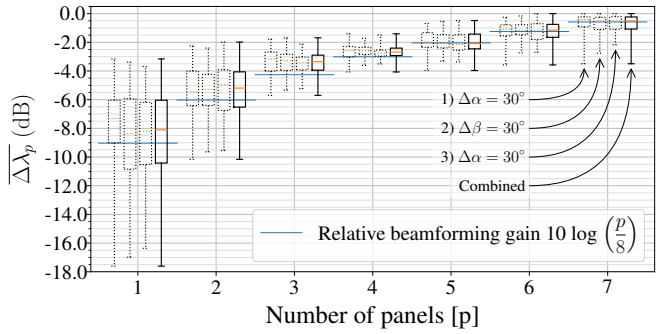}}
    \caption{Gain ratio per mobility pattern (dotted) and combined (full line). Median values are depicted in orange.}
    \label{fig:general_gain}
    \vspacebelowfig
\end{figure}

\Cref{fig:general_gain} plots the \gls{de} gain trade-off, dependent on the number of antenna panels. The median gain trade-off, across all channel snapshots (full lines), increases logarithmically with the number of employed panels; however, the observed gain only starts conforming with the beamforming gain for an \gls{hmd} with 5 or more panels. This leads us to two conclusions 1) the beamforming gain for 1--4 panels is overestimated since not all elements on an 8-panel \gls{hmd} are illuminated, and 2) 4~panels capture the available diversity gain (sufficient azimuth coverage) and adding any additional panels results in merely beamforming gain. We also observe a reduction in the spread of the distributions as the number of employed panels grows, with the most noticeable improvements from 1~to~2 and from 2~to~3~panels, while 4~panels offer a marginal improvement over 3~panels. Finally, a 4-panel \gls{hmd} follows the gain trend of an 8-panel \gls{hmd} the most accurately due to its good symmetry (see \Cref{fig:panel_configurations}), i.e., an evenly distributed azimuth gain. 

The dotted distributions in \Cref{fig:general_gain} show that all, except a 2-panel \gls{hmd}, experience somewhat better performance during the initial $\Delta\alpha=30^{\circ}$ rotation, when the \gls{ue} is standing upright. This is most visible for a single-panel \gls{hmd}. We assume the carpeted floor and numerous table and chair legs result in poor \gls{mpc} reflection, that the forward-pitched \gls{ue} could receive. A 2-panel \gls{hmd} exhibits favorable performance in the final part of the mobility. This is due to the \gls{nlos} experiments, where the second $\Delta\alpha=30^{\circ}$ rotation, at some point, directly exposes one of the two panels to the \gls{los} that bypasses the phantom.

% \subsection{Gain spread and persistence}
\subsection{Gain volatility}
\begin{figure}[h]
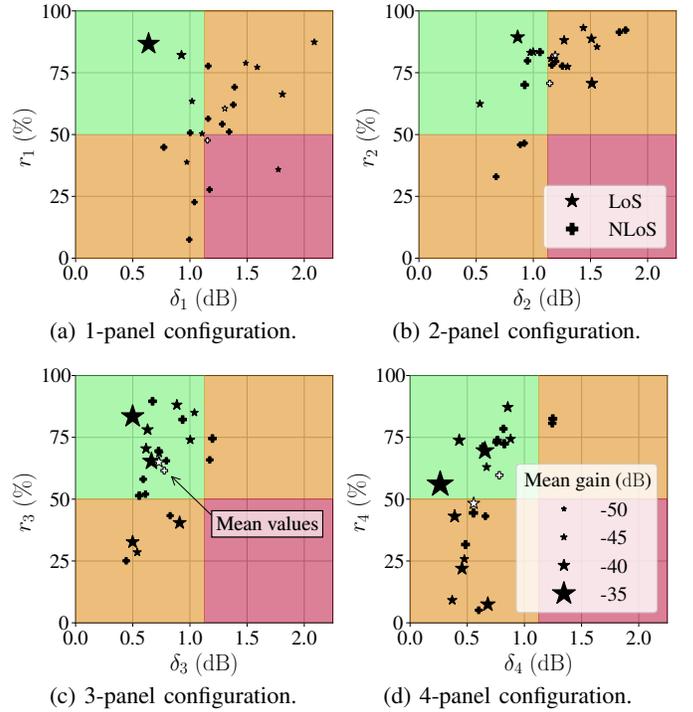

     \centering     
     \begin{subfigure}[b]{0.235\textwidth}
         \centering
         \resizebox{1\columnwidth}{!}{\input{figure-conference/volatility_00000010.pgf}}
         \vspace{-0.5 cm}
         \caption{1-panel configuration.}
     \end{subfigure}
     \hfill
     \begin{subfigure}[b]{0.235\textwidth}
         \centering
         \resizebox{1\columnwidth}{!}{\input{figure-conference/volatility_10001000.pgf}}
         \vspace{-0.5 cm}
         \caption{2-panel configuration.}
     \end{subfigure}
     \hfill
     \begin{subfigure}[b]{0.235\textwidth}
         % \vspace{0.3 cm}
         \centering
         \resizebox{1\columnwidth}{!}{\input{figure-conference/volatility_01010010.pgf}}
         \vspace{-0.5 cm}
         \caption{3-panel configuration.}
     \end{subfigure}
     \hfill
     \begin{subfigure}[b]{0.235\textwidth}
         % \vspace{0.2 cm}
         \centering
         \resizebox{1\columnwidth}{!}{\input{figure-conference/volatility_01010101.pgf}}
         \vspace{-0.5 cm}
         \caption{4-panel configuration.}
     \end{subfigure}
     \hfill     
    \vspace{-0.3 cm}
    % \caption{\gls{de} gain standard deviation ($\delta$) and autocorrelation ($r$). Least to most prosperous regions (approx.): red (bottom right), orange (bottom left, top right), and green (top left). Marker size shows the mean gain during a measurement.}
    \caption{\gls{de} gain standard deviation ($\delta$) and autocorrelation ($r$). Least to most prosperous regions (approx.): red, orange, and green (darkest to brightest in grayscale). Marker size shows the mean gain during a measurement.}
    \label{fig:volatility}
    \vspacebelowfig
\end{figure}

The gain's standard deviation ($\delta_p$) and autocorrelation ($r_p$) are depicted in
\Cref{fig:volatility}. The \gls{los} and \gls{nlos} scenarios are separated by star/cross markers, while marker size shows the average gain for each of the 22 measurements. A single panel yields high gain volatility, since adjacent measurements (less than 1~cm apart) may exhibit a high difference in gain as the panel moves or rotates from \gls{los} to \gls{nlos} or vice-versa. Gain spread remains high for a 2-panel \gls{hmd}, however, gain perturbations are gradual over time (high autocorrelation). This benefits the \gls{hmd} since resolution adaptations and handovers are less frequent. 
% \glspl{hmd} with 3 or more panels can still show low autocorrelation, however, they are less likely to experience outage due to less pronounced gain spread. 
An \gls{hmd} with 3 or more panels shows less pronounced gain spread and higher mean gain. The low autocorrelation points originate from more adverse circumstances, where 3 or 4 panels receive the strongest \gls{mpc} from a large angle, relative to the individual radiation patterns. 
For example, take a forward-pitched 4-panel \gls{hmd} at positions UE$_4$ or UE$_9$, where the \gls{los} angle of arrival strides around $\pm$\SI{45}{\degree} or right in\nobreakdash-between two panels. Yet, this is a less severe problem since fluctuations around a high mean gain are less likely to cause service outage. 
Conversely, a 2-panel \gls{hmd} would have one of its panels well-exposed in this case, while it could undergo longer faded periods in other scenarios. Hence, the high autocorrelation and standard deviation.
Gain volatility does not significantly change for 5--8 panels.

\subsection{Minimal service and channel capacity trade-off}

\Cref{tab:minimal_service} shows that the reduction in the minimal attainable service level is most pronounced when moving from 3~to~2 and from 2~to~1~panel/\nobreakdash-s, which results in \SI{0.58}{} and \SI[per-mode=repeated-symbol]{1.35}{\bit\per\second\per\hertz} lower channel capacity, respectively. The noticeably smaller differences between other configurations show no clear trend. 

\begin{table}[h]
    \centering
    \vspace{-3mm}
    \caption{Minimal service trade-off (\SI{97}{\percent} reliability). }
    \vspace{-1mm}
    \begin{tabular}{l|c|c|c|c|c|c|c}
        \toprule
        Number of panels & 1 & 2 & 3 & 4 & 5 & 6 & 7 \\
        % \textbf{Number of panels} & \textbf{1} & \textbf{2} & \textbf{3} & \textbf{4} & \textbf{5} & \textbf{6} & \textbf{7} \\
        \midrule
        % $\Delta C_{p,97}$ & -3.07 & -1.72 & -1.14 & -0.86 & -0.71 & -0.47 & -0.31 \\
        $\Delta C_{p,97}$ [\SI[per-mode=repeated-symbol]{}{\bit\per\second\per\hertz}] & 3.07 & 1.72 & 1.14 & 0.86 & 0.71 & 0.47 & 0.31 \\
        $\Delta C_{p+1,97} - \Delta C_{p,97}$ & 1.35 & 0.58 & 0.28 & 0.15 & 0.24 & 0.16 & 0.31 \\
        \bottomrule
    \end{tabular}
    \label{tab:minimal_service}
\end{table}

\begin{figure}[t]
    \vspace{3mm}
    \centering
    \resizebox{0.98\columnwidth}{!}{\input{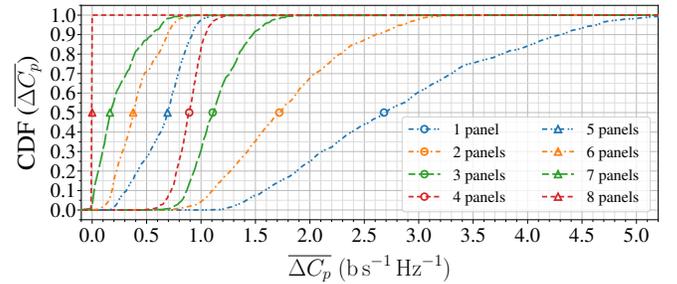}}
    \caption{Channel capacity trade-off.}
    \label{fig:capacity}
    \vspace{-6mm}
\end{figure}

\Cref{fig:capacity} confirms the poor performance for either 2~panels or a single panel configuration, which can suffer in excess of \SI{3}{} and \SI[per-mode=repeated-symbol]{5}{\bit\per\second\per\hertz} lower channel capacity than an 8-panel \gls{hmd}, correspondingly. However, we also notice that a 3\nobreakdash-panel \gls{hmd} exhibits a more slanted CDF than the 4\nobreakdash-panel \gls{hmd}, in addition to achieving \SI[per-mode=repeated-symbol]{0.7}{\bit\per\second\per\hertz} lower capacity in the worst case. A 4\nobreakdash-panel \gls{hmd} offers the most deterministic performance, since its CDF has the steepest slope, best matching 8\nobreakdash-panel performance. We can conclude, that, contrary to the previous two performance metrics, the 4\nobreakdash-panel configuration has a slight edge over the 3\nobreakdash-panel \gls{hmd} since the performance gap starts to increase already when decreasing from 4~to~3~panels.

\subsection{Profit from a rear headband}
\label{sec:rear_headband_results}
\begin{figure}[h]
    \centering
    \resizebox{0.98\columnwidth}{!}{\input{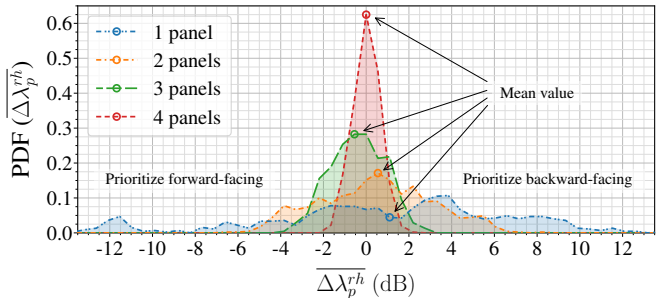}}
    % \caption{Potential gain of \glspl{hmd} sporting a rear headband.}
    % \caption{\gls{de} gain increase for backward-facing configuration.}
    \caption{\Gls{pdf} of the gain ratio between the backward- and forward-facing panel configuration.}
    \label{fig:headband_gain}
    \vspacebelowfig
\end{figure}

\Cref{fig:headband_gain} shows that, across all available snapshots, a backward-facing single-panel setup yields higher mean gain, which means that orienting antennas towards the ceiling results in higher gain than pointing them towards the floor, since the \gls{hmd} spends the majority of time tilted downwards. As noted in \Cref{sec:results_general}, the reason for the lower gain might be a carpeted and cluttered floor. A 2-panel setup also mildly prioritizes the backward-facing configuration, however, we associate this with our mobility pattern design, where the backward-facing configuration prospers when pitching towards or away from the \gls{ap} (\SI{15}{} of the total \SI{33}{\second}). Conversely, a 3-panel \gls{hmd} performs better in the forward-facing configuration, where 2 panels are placed diagonally at the back and only one panel at the front, thus, emphasizing ceiling-bound \glspl{mpc}. The remaining panel configurations do not favor a forward- or a backward facing configuration, so we only include the representative 4-panel \gls{hmd} results.

\section{Conclusion}
\label{sec:conclusion}

The work at hand evaluates the performance of multiple \gls{mmw} antenna array configurations an \gls{xr} \gls{hmd}, based on indoor channel sounding data. 
We introduce physical layer performance metrics, relevant for selecting the optimal \gls{mmw} array configuration on an \gls{hmd}, and use them to evaluate the performance of an \gls{hmd} employing 1--7 antenna panels.
Results demonstrate that fitting an \gls{hmd} with a single panel will offer poor performance in all aspects. Using an additional panel improves temporal gain stability (few abrupt changes), yet, a 2\nobreakdash-panel \gls{hmd} still struggles in providing sufficient gain during rotation. At least 3 panels are needed to offer a stable and consistent gain, while including a 4$^{th}$ panel further improves the minimal achievable service quality and channel capacity by a limited amount. Any additional patch antenna arrays will offer minute performance improvements at the cost of greater \gls{hmd} complexity and higher energy consumption. 
% We conclude, that an \gls{hmd} with \gls{mmw} support should feature at least 3 directional patch antenna arrays -- e.g., two behind the ears, providing side and back coverage, and one next to the lenses/display, illuminating the front. 
In future work, we aim to closely investigate the effects of blockage and derive the number of antenna arrays that are required for maximally benefiting from the available \glspl{mpc} and mitigating the effects of blockage. 
% Moreover, future work should evaluate the most prosperous array configurations (2--4 panels) from the perspective of multi-stream \gls{mimo} transmission to provide accurate upper performance bounds, and extend the list of relevant performance metrics. 
Moreover, future work should evaluate the most prosperous array configurations (2--4 panels) from the perspective of multi\nobreakdash-stream \gls{mimo} transmission to provide accurate upper performance bounds and evaluate the role of the number of antennas (in addition to arrays) on small\nobreakdash-scale fading. 
Looking from another perspective, the results also support the advocated trend for 6G networks: distributed infrastructure will benefit reliability and robustness. 
% Thus, multi-array \glspl{ap} and multi-\gls{ap} deployments are a worthy subject for future wireless \gls{hmd} research.
Thus, the potential of lowering \gls{hmd} complexity by relying on a distributed antenna infrastructure is a worthy subject for future research.
% For example, the gain distribution across channel eigenmodes.
% For example, by evaluating eigenmode distribution and investigating the channel condition number (multi-user \gls{mimo}). 
% Moreover, evaluating the benefits of massive \gls{mimo} versus using a limited number of \gls{mmw} antennas per panel would provide \gls{xr} \gls{hmd} designers with hardware form factor requirements, while assessing the performance for multi-user \gls{mimo} would give rough estimates of how many users an \gls{ap} can serve w.r.t. physical channel limitations. 
% Moreover, future work should parameterize the small-scale \gls{mmw} channel behavior during mobility, such as the one described in this work, and extend existing channel models to cover \gls{xr} use cases.

% Evaluate redundancy for reliability -- what is the minimum capacity that we can achieve for duplicating the mainstream over 2, 3, etc. signal paths or pre-coding using multiple resolution classes?

\section*{Acknowledgement}

\noindent
\footnotesize{Thanks to Meifang Zhu, Gilles Callebaut, and the volunteers at Lund Uni.}

\vspace{-1 mm}
\noindent
\begin{minipage}{0.18\columnwidth}
    \vspace{1 mm}
    \includegraphics[width=\columnwidth]{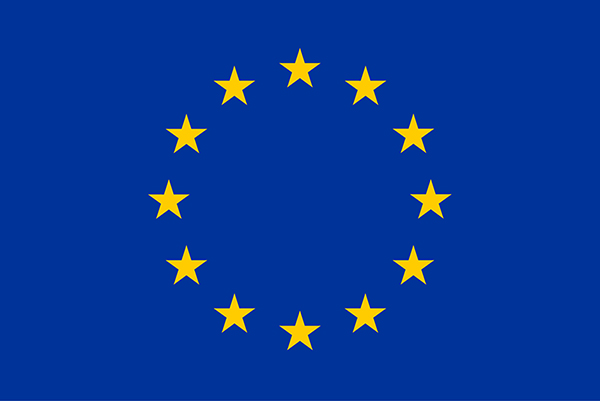}
\end{minipage} \hfill
\begin{minipage}{0.803\columnwidth}
    \vspace{1 mm}
    % \footnotesize{This work has received funding from the European Union’s Horizon 2020 research and innovation program under grant agreement No. 861222 (MINTS).}
    \footnotesize{This work has received funding from the EU’s Horizon 2020 and Horizon Europe programmes under grant agreements No. 861222 (MINTS) and 101096302 (6G Tandem).}
\end{minipage}

\noindent
\footnotesize{The work is also partially supported by the Horizon Europe Framework Programme under the
Marie Skłodowska-Curie grant agreement No.,101059091, the Swedish Research Council
(Grant No. 2022-04691), the strategic research area ELLIIT, Excellence Center at Linköping
— Lund in Information Technology, and Ericsson. }

\bibliographystyle{ieeetr}
\bibliography{refs}

\begin{thebibliography}{10}

\bibitem{FIROUZI2022101840}
F.~Firouzi {\em et~al.}, ``{The Convergence and Interplay of Edge, Fog, and
  Cloud in the AI-Driven Internet of Things (IoT)},'' {\em Information
  Systems}, vol.~107, p.~101840, 2022.

\bibitem{shafi_5g_2017}
M.~Shafi {\em et~al.}, ``{5G}: {A} {Tutorial} {Overview} of {Standards},
  {Trials}, {Challenges}, {Deployment}, and {Practice},'' {\em IEEE JSAC},
  vol.~35, pp.~1201--1221, June 2017.

\bibitem{zhou_ieee_2018}
P.~Zhou {\em et~al.}, ``{IEEE} 802.11ay-{Based} {mmWave} {WLANs}: {Design}
  {Challenges} and {Solutions},'' {\em IEEE Communications Surveys \&
  Tutorials}, vol.~20, no.~3, pp.~1654--1681, 2018.

\bibitem{THzMagazine}
X.~Cai {\em et~al.}, ``{Toward 6G with Terahertz Communications: Understanding
  the Propagation Channels},'' {\em IEEE Commun. Mag.}, to appear, 2023.

\bibitem{cai_dynamic_2020}
X.~Cai {\em et~al.}, ``Dynamic {Channel} {Modeling} for {Indoor}
  {Millimeter}-{Wave} {Propagation} {Channels} {Based} on {Measurements},''
  {\em {IEEE} Trans. Commun.}, vol.~68, pp.~5878--5891, Sept. 2020.

\bibitem{choi_measurement_2018}
T.~Choi {\em et~al.}, ``Measurement {Based} {Directional} {Modeling} of
  {Dynamic} {Human} {Body} {Shadowing} at 28 {GHz},'' in {\em {2018 GLOBECOM}},
  (Abu Dhabi, United Arab Emirates), pp.~1--6, IEEE, Dec. 2018.

\bibitem{hoque_distinguishing_2014}
R.~Hoque {\em et~al.}, ``{Distinguishing Performance of 60-{GHz} Microstrip
  Patch Antenna for Different Dielectric Materials},'' in {\em 2014 ICEEICT},
  (Dhaka, Bangladesh), pp.~1--4, IEEE, Apr. 2014.

\bibitem{gustafson_characterization_2012}
C.~Gustafson and F.~Tufvesson, ``{Characterization of 60 GHz Shadowing by Human
  Bodies and Simple Phantoms},'' in {\em 2012 {EUCAP}}, (Prague, Czech
  Republic), pp.~473--477, IEEE, Mar. 2012.

\bibitem{vaha-savo_empirical_2020}
L.~Vähä-Savo {\em et~al.}, ``Empirical {Evaluation} of a 28 {GHz} {Antenna}
  {Array} on a {5G} {Mobile} {Phone} {Using} a {Body} {Phantom},'' {\em {IEEE}
  Trans. Antennas Propag.}, 2020.

\bibitem{hejselbak_measured_2017}
J.~Hejselbak {\em et~al.}, ``Measured 21.5 {GHz} {Indoor} {Channels} {With}
  {User}-{Held} {Handset} {Antenna} {Array},'' {\em {IEEE} Trans. Antennas
  Propag.}, vol.~65, pp.~6574--6583, Dec. 2017.

\bibitem{aggarwal_first_2019}
S.~Aggarwal {\em et~al.}, ``A {First} {Look} at 802.11ad {Performance} on a
  {Smartphone},'' in {\em {Proc. ACM mmNets}}, pp.~13--18, Oct. 2019.

\bibitem{zhao_user_2017}
K.~Zhao {\em et~al.}, ``User {Body} {Effect} on {Phased} {Array} in {User}
  {Equipment} for the {5G} {mmWave} {Communication} {System},'' {\em {IEEE}
  Antennas Wireless Propag. Lett.}, vol.~16, pp.~864--867, 2017.

\bibitem{xu_radiation_2018}
B.~Xu {\em et~al.}, ``Radiation {Performance} {Analysis} of 28 {GHz} {Antennas}
  {Integrated} in {5G} {Mobile} {Terminal} {Housing},'' {\em IEEE Access},
  vol.~6, pp.~48088--48101, 2018.

\bibitem{gomes_mm_2018}
R.~Gomes {\em et~al.}, ``{A mmWave Solution to Provide Wireless Augmented
  Reality in Classrooms},'' in {\em {2018 ISWCS}}, (Lisbon), pp.~1--6, IEEE,
  Aug. 2018.

\bibitem{struye_opportunities_2022}
J.~Struye {\em et~al.}, ``Opportunities and {Challenges} for {Virtual}
  {Reality} {Streaming} over {Millimeter}-{Wave}: {An} {Experimental}
  {Analysis},'' in {\em 2022 {NoF}}, (Ghent, Belgium), pp.~1--5, IEEE, Oct.
  2022.

\bibitem{sounder_paper}
X.~Cai {\em et~al.}, ``{A Switched Array Sounder for Dynamic Millimeter-Wave
  Channel Characterization: Design, Implementation and Measurements},'' {\em
  {IEEE} Trans. Antennas Propag.}, submitted, 2023.

\bibitem{kunin_rotation_2007}
M.~Kunin {\em et~al.}, ``Rotation {Axes} of the {Head} {During} {Positioning},
  {Head} {Shaking}, and {Locomotion},'' {\em Journal of Neurophysiology},
  vol.~98, pp.~3095--3108, Nov. 2007.

\bibitem{molisch_wireless_book}
A.~F. Molisch, {\em Wireless Communications}, p.~398.
\newblock Wiley Publishing, New Jersey, US, 3rd~ed., 2022.

\bibitem{qi_energy_2022}
X.~Qi {\em et~al.}, ``{Energy-Efficient Power Allocation in Multi-User
  mmWave-NOMA Systems With Finite Resolution Analog Precoding},'' {\em {IEEE}
  Trans. Veh. Commun}, vol.~71, no.~4, pp.~3750--3759, 2022.

\bibitem{chiani_distribution}
M.~Chiani, ``{Distribution of the Largest Eigenvalue for Real Wishart and
  Gaussian Random Matrices and a Simple Approximation for the Tracy-Widom
  Distribution},'' {\em CoRR}, vol.~abs/1209.3394, 2012.

\bibitem{nightingale_subjective_2013}
J.~Nightingale {\em et~al.}, ``{Subjective Evaluation of the Effects of Packet
  Loss on HEVC Encoded Video Streams},'' in {\em {2013 ICCE}}, (Berlin,
  Germany), pp.~358--359, IEEE, Sept. 2013.

\end{thebibliography}

\end{document}